\documentclass{article}
\usepackage[T1]{fontenc} 
\usepackage[utf8]{inputenc} 
\usepackage{hyperref}

\usepackage{graphicx}
\usepackage{subcaption} 
\usepackage{float}
\usepackage{subcaption}

\usepackage{graphicx} \usepackage[margin=1in]{geometry}
\usepackage{
    amsmath,
    latexsym,
    color,
    multirow,
    array,
    listings,
    enumitem,
    multicol,
    balance,
    graphicx,
    mathrsfs
}
\usepackage[dvipsnames]{xcolor}
\usepackage{cite}
\usepackage{todonotes}

\newcommand{\etal}{{\it et al.~{}}}

\def\blackslug{\vrule height6pt width3pt depth1pt}
\newtheorem{auxdefn}{Definition}[section]

\title{Homoglyph-based Adversarial Perturbation of Introductory Computer Science Theory Problems}
\author{
Aidan Alexander\textsuperscript{*} \and
Chitrangada Juneja\textsuperscript{*} \and
Napaluck Tontrasathien \and 
Miro Vanek \and
Reyan Ahmed \and
Saumya Debray \and
Sazzadur Rahaman
}

\date{\today}

\begin{document}

\maketitle

\def\thefootnote{*}\footnotetext{These authors contributed equally to this work}\def\thefootnote{\arabic{footnote}}
\begin{abstract}
    Different AI tools such as ChatGPT, Gemini, and Claude are becoming very popular. Although they are helpful for many day-to-day tasks, they can be used in unexpected ways. For example, the learning objectives of a course may not be achieved if students use these tools to solve their homework problems. This paper proposes a simple method to address this issue in the lazy student model. The method uses homoglyph-based adversarial perturbation to first modify the question without changing the semantic meaning of the question. Then a few characters are perturbed by their homoglyphs. Our experimental result shows the theoretical problems of introductory computer science courses can be effectively perturbed. We also propose an interactive tool to conveniently use our method.
\end{abstract}
\section{Introduction}\label{sec:introduction}


AI-based tools such as ChatGPT have an impressive ability to perform language-based tasks such as writing text, answering questions, and generating code.  Although useful in a broader social context~\cite{fang2024large,colelough2025neuro}, this is a significant problem in academic settings, since students can use such tools to do assignments and solve homework problems without necessarily learning or understanding the material.  For example, many of our colleagues report a sharp spike in AI-based cheating in lower-division Computer Science courses at our institution, and several have been forced to restructure course assessments as a result.  Scholarly research on the topic, as well as anecdotal evidence in academic online discussion forums, suggests that the problem is widespread across institutions and disciplines \cite{chen2024plagiarism,pudasaini2025survey,adnan2025cheating}.

There are generally three broad approaches to dealing with AI-assisted submissions by students: (1) restructuring the course to \emph{allow} AI-generated content; (2) restructuring the course to \emph{sidestep} the issue, e.g., by deemphasizing homework assignments (which are more susceptible to AI-assisted cheating) in favor of in-class work; and (3) \emph{disallowing} AI-generated content.
We do not take a position on whether one approach is better than another: 
this depends on the learning objectives of the class and how the use of AI tools impacts those objectives.    
In any case, we feel that it is important for instructors to have good options for whichever approach they feel is the most appropriate for their class.  This paper focuses on broadening the options available for the third approach, where AI-generated content is disallowed.

In classes where students are not allowed to submit AI-generated content, enforcement usually involves analyzing student submissions for indicators of AI-generated content and penalizing submissions where such indicators are found.  
This approach is problematic for many reasons, among them that identifying AI-generated content is unreliable and error-prone, and mistakenly accusing anyone of cheating can be traumatic.  We believe that a better approach is to deter AI-based cheating in the first place.  To this end, we investigate an approach where we modify the assignment prompts in such a way that (1) they are visually similar to the original prompts; but (2) they cause AI-based tools to provide incorrect answers. We refer to this approach as \emph{adversarial perturbation} of the assignment prompts.
Our goal in doing this is to reduce the usefulness of AI-based tools in solving homework problems and thus deter students from using them in the first place.

Homoglyph-based attacks have long been used against detectors that scan text for undesirable content, e.g., spam filters and phishing detectors \cite{liu2007fighting,sokolov2020visual,gabrilovich2002homograph}.
More recently, they have been used to evade hate-speech filters on social media platforms \cite{cooper2023hiding} and plagiarism detectors aimed at identifying AI-generated text \cite{creo2025silverspeak,wolff2020attacking}. Huang \etal~\cite{huang2025math} show that LLMs’ strong math performance often relies on pattern memorization rather than true reasoning, as their accuracy drops significantly when problems are fundamentally altered through hard perturbations. Hao \etal~\cite{hao2025investigation} introduce a new benchmark (PutnamGAP) using mathematically equivalent but varied problems to show that LLMs’ reasoning performance drops significantly under non-mathematical changes, revealing limited robustness and generalization.
Recent work by Salim \etal \cite{salim2024impeding} on introductory programming problems indicates that, when faced with problems that AI tools do not answer correctly, students end up doing the work themselves.  Our goal here is similar, albeit for a very different problem domain: computer science theory problems, which are generally mathematical in nature and are typically shorter than programming problems and thus tend to have fewer opportunities for adversarial perturbation.



\begin{figure}[h]
    \centering
    \includegraphics[width=0.5\textwidth]{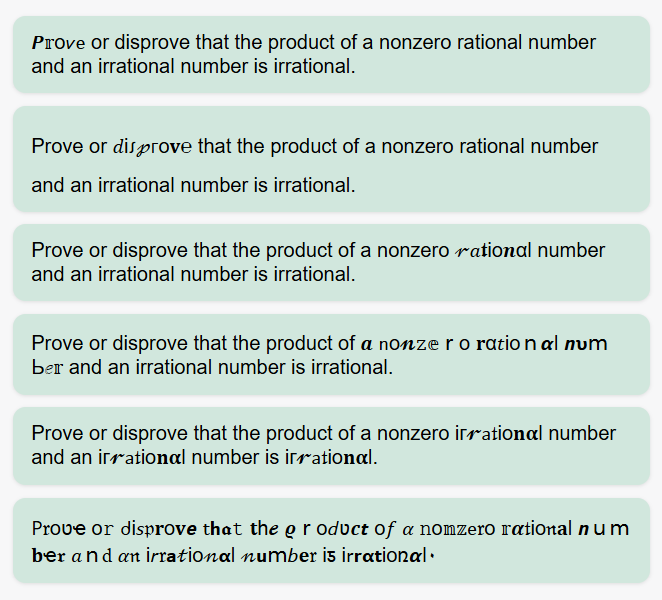}
    \caption{Despite the fact that several parts of the question are perturbed, as shown in this figure, most popular LLMs, including ChatGPT and Gemini, can answer the question without any difficulty. This observation indicates the robustness of modern LLMs.}
    \label{fig:rational_prompt}
\end{figure}
\begin{figure}[h]
    \centering
    \includegraphics[width=0.95\textwidth]{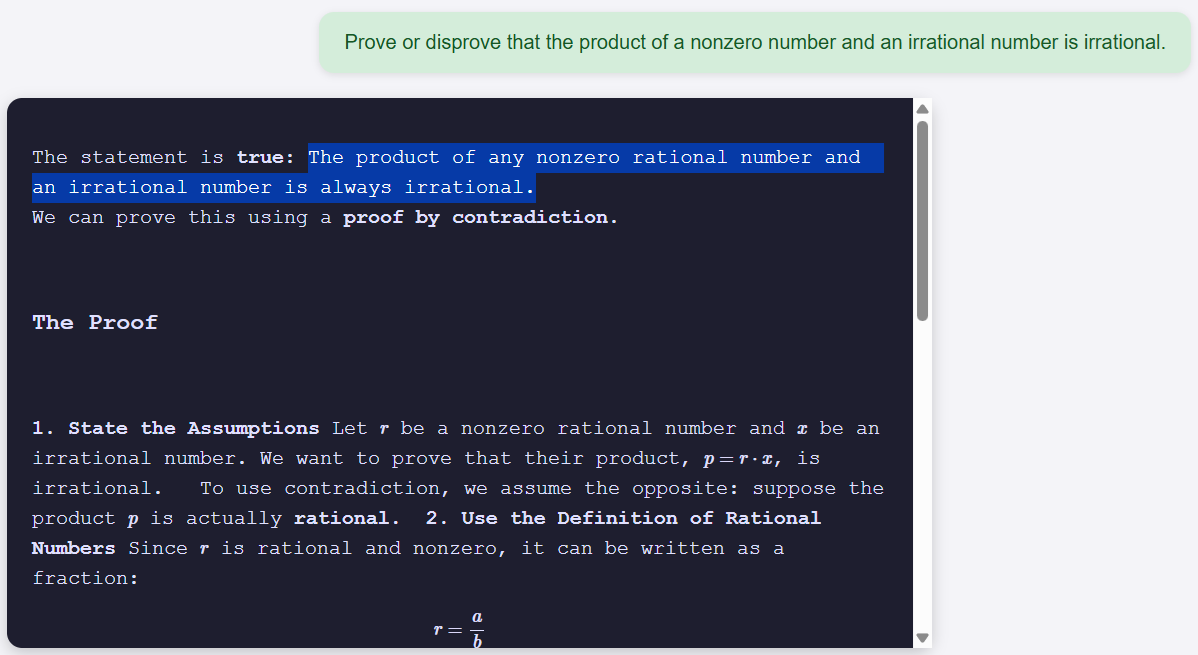}
    \caption{A common type of response from popular LLM models, including ChatGPT and Gemini, indicating the biases of the models towards the pretrained dataset. An important adjective ``rational'' is missing in the prompt and the model easily adds back the missing word since it is familiar with the question from the training dataset. This observation is later used in our method.}
    \label{fig:rational_chat}
\end{figure}

On the face of it, the idea seems fairly straightforward: replace one or more characters in the problem prompt with appropriate homoglyphs and check whether the result is effective in fooling the Large Language Models (LLM), repeating the process if necessary.  Unfortunately, modern LLMs are powerful enough that this straightforward approach is not very effective.  The problem arises from the fact that theory problems assigned for homeworks are often taken from textbooks or other similar sources that, because of their wide availability, are well represented in LLM training data and thus have been ``memorized'' by the LLMs.  As a result, LLMs are able, in most cases, to undo the effects of homoglyph-based perturbations to problem prompts.  In fact, modern LLMs can often repair prompts even when entire words or phrases have been perturbed or deleted, see Figures~\ref{fig:rational_prompt} and \ref{fig:rational_chat}. To get around this problem we use the following approach.  We first identify key words or symbols in the prompt that are indicative of memorization by the LLM.  We then change some of those key words or symbols such that the resulting problem, which is now no longer the same as the original problem, nevertheless tests the same concepts as the original prompt.  Finally, we use homoglyph-based perturbation on the modified problem prompt so as to hinder proper tokenization by the LLM front ends.

Our method is simple and effective: it can fool all the popular LLM models. We have taken a set of homework problems from a well-known discrete mathematics book~\cite{discrete2007mathematics}. Our experimental result shows that all the questions can be perturbed easily by only applying a few perturbations. We also propose an interactive tool to apply our method.
\section{Background}\label{sec:background}


In this section we mention the assumptions we made and a brief description of LLMs.

\subsection{Assumptions}\label{sec:assumptions}

\begin{enumerate}
    \item No assumptions about which LLMs a student may use $\Rightarrow$ need to be effective against a range of different LLMs.
    \item {\em Lazy student model}: assume that students copy-paste the assignment prompt to the LLMs input.  This is consistent with other work in this area~\cite{santos2026llm,puthumanaillam2025lazy}.
\end{enumerate}

\subsection{LLM Architecture}


Large language models are typically trained in a multi-stage pipeline, beginning with a large-scale \textit{pretraining phase}. During pretraining, the model is exposed to massive corpora of text and learns statistical patterns of language through a self-supervised objective; most commonly next-token prediction. At each step, the model is given a sequence of tokens and trained to predict the most likely subsequent token. Over time, this objective leads the model to internalize grammar, factual associations, reasoning patterns, and even some world knowledge. Crucially, this learning does not rely on explicit labels; instead, the structure inherent in natural language provides the training signal. The scale of both the dataset and the model parameters plays a central role in determining capability, with larger models generally able to capture more nuanced patterns.

Before any learning can occur, raw text must go through \textit{tokenization}, which converts continuous text into discrete units (tokens). These tokens might correspond to words, subwords, or even individual characters, depending on the tokenization scheme (e.g., byte-pair encoding or unigram language models). The choice of tokenizer has significant implications: it affects vocabulary size, efficiency, and how well rare or complex words are represented. Once tokenized, each token is mapped to a dense vector representation known as an \textit{embedding}. These embeddings are learned during training and serve as the model’s interface between discrete symbols and continuous computation. They encode semantic and syntactic relationships such that tokens with similar meanings or usage patterns occupy nearby regions in the embedding space.

After pretraining, models typically undergo a \textit{post-training} phase to better align their outputs with human expectations. This stage may include supervised fine-tuning on curated datasets, where the model learns to follow instructions or produce more contextually appropriate responses. It is often followed by alignment techniques such as reinforcement learning from human feedback (RLHF), where human evaluators rank outputs and the model is optimized to prefer higher-quality responses. Additional methods, like constitutional training or preference optimization, may also be used to improve safety, factuality, and usefulness. Together, these post-training steps refine the raw generative capabilities learned during pretraining into behavior that is more controlled, reliable, and aligned with real-world applications.

Recent advancements in LLMs have produced some of the strongest models that perform very well in problem solving. Not only are these models highly capable of solving problems, but they also appear \textit{initially} to be quite robust to different types of attacks. Our initial experiments have shown that even when we perturb many significant characters, the models still perform very well.

For example, consider the question: ``Prove or disprove that the product of a nonzero rational number and an irrational number is irrational.'' This is a classical question in discrete mathematics, and if we ask this question to models like ChatGPT or Gemini, they can easily answer it. Even if we perturb several parts of the question and present it to an LLM (see Figure~\ref{fig:rational_prompt}), the models are still able to answer it correctly with ease. The last perturbation shown in the figure modifies nearly all characters, making the question difficult to read even for humans, although the models can still answer it correctly. This demonstrates the robustness of these models.
In this paper we present a method where despite such a high robustness of these models, we can still fool them just perturbing a very few number of characters that is not noticeable to human eyes.

\begin{figure}[h]
    \centering
    \includegraphics[width=0.5\textwidth]{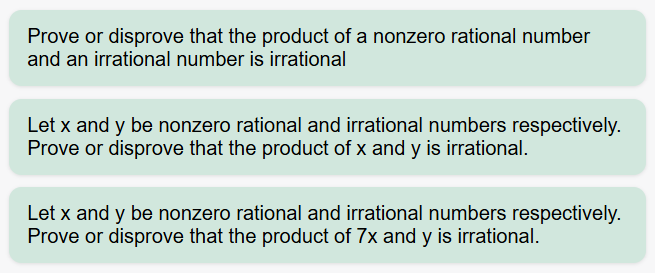}
    \caption{Illustrating possible modifications of a question without changing the semantic meaning. The first prompt is the original questions. The variables $x$ and $y$ are introduced in the second prompt without changing the semantic meaning. The third prompt is slightly different: $7x$ is used instead of $x$.}
    \label{fig:rational_prompt2}
\end{figure}

\section{Proposed Method}\label{sec:research}

Our method, if needed after necessary modification to the question without changing the main goal, effectively perturbs the question just using a few number of characters. The method deliberately disrupts the tokenization process using a set of techniques described below, causing characters or words to be skipped or misinterpreted by the tokenization phase and changing the meaning of the question. As a result, the model’s internal representations are corrupted, leading to incorrect outputs. This demonstrates that perturbations at the tokenization and embedding stage can have a significant impact on the behavior and reliability of LLMs. We now describe the main steps of our method.

\subsection{Step 1: Find out a Part for Perturbation}

Despite the fact that the modern models are very robust, we will propose a method that can effectively fool the models. The observation that led us to develop this method is realizing that even if we remove important parts from the question, still the models can answer the questions! There might be multiple reasons behind this: one possible thing might be that the models are just memorizing the questions since we have tried some famous questions of discrete mathematics. For example consider the questions ``Prove or disprove that the product of a nonzero rational number and an irrational number is irrational.'' This is a question that can be found in popular discrete mathematics books as well in many online tutorials. Since the models are pretrained from datasets that contain these resources, the model might be just memorizing them. 

Another related explanation is that the models use an algorithm to fill up the missing information. And the algorithm is biased towards the dataset that was used to pretrain. Such an algorithm might be necessary for real models since the user can unintentionally miss to provide some information or to save the systems for malicious attacks where intentionally the question is modified. And this is also consistent with the fact that even if the whole question is perturbed, the model can regenerate the original question as illustrated in Figure~\ref{fig:rational_prompt}. Some perturbation might be effective, but the model just replaced those perturbations using some algorithm to find the original question.

Independent of the method the models use to fill in the missing information, we observed that the models are biased towards some specific parts of the original question that were not perturbed. For example, in the question, the adjective ``rational'' of ``a nonzero rational number'' is really important. Surprisingly, if we delete this adjective and ask ``Prove or disprove that the product of a nonzero number and an irrational number is irrational.'', the models without hesitation replaces ``nonzero number'' by ``nonzero rational number''; which is a clear indication that the models are biased towards using ``nonzero rational number''. That is the first step of our method: we want to find out a part of the original unperturbed question that the model is biased to and even that part will be missing in the question the models are going to replace it by adding the missing part that the model is biased to. In the later steps of our method we will effectively perturb that part of the question using the model's biasness.

\begin{figure}[h]
    \centering
    \includegraphics[width=0.5\textwidth]{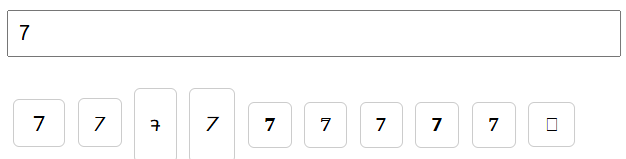}
    \caption{Illustrating different homoglyphs of the numeric character ``7''. Some homoglyphs are readable by humans and others are not.}
    \label{fig:homoglyphs7}
\end{figure}

\subsection{Step 2: Modify the Question If Needed}

Our next step is to modify the questions with some simple changes if needed (see Figure~\ref{fig:rational_prompt2}). If the question already has numeric characters or mathematical expressions, then most of the time this step is not required. For example, consider another classical question ``Prove or disprove that if you have an 8-gallon jug of water and two empty jugs with capacities of 5 gallons and 3 gallons, respectively , then you can measure 4 gallons by successively pouring some of or all of the water in a jug into another jug.'' In this question we already have a couple of numeric characters, e.g. ``8'', ``5'', and ``3''. On the other hand, the question ``Prove or disprove that the product of a nonzero rational number and an irrational number is irrational'' does not have any numeric characters or mathematical expression. However, since we are considering problems of discrete mathematics it is easy to modify the question by introducing some numeric characters or mathematical expressions without changing the semantic meaning of the question. This is in general true for many problem solving questions. We can modify the particular question about irrational numbers to ``Let $x$ and $y$ be nonzero rational and irrational numbers respectively. Prove or disprove that the product of $x$ and $y$ is irrational.'' We can also add one or more coefficients without changing the question significantly, e.g. ``Let $x$ and $y$ be nonzero rational and irrational numbers respectively. Prove or disprove that the product of $7x$ and $y$ is irrational.'' In the second modification we added a coefficient $7$ to $x$, however, if a student has the problem solving skill to answer the question without the modification, the student must be able to answer the modified one. The reason for these modifications will become clear in the following sections, in brief, it will help us to generate questions that the models will be unable to solve.

\subsection{Step 3: Find Effective Homoglyphs}

There are several \href{https://github.com/codebox/homoglyph?tab=readme-ov-file}{homoglyphs} that can be applied to perturb the questions. Each character has multiple homoglyphs, and we can replace a character that appears in the question by a homoglyph of the character. For example, the numeric character ``7'' has multiple homoglyphs (see Figure~\ref{fig:homoglyphs7}). However, not all homoglyphs are effective, specially most popular LLMs can figure out that a homoglyph is used. Hence the LLMs will replace the homoglyph by the original character. However, there exists homoglyphs that the LLMs can not understand. Also, there are homoglyphs that look quite different from the original character, see the third homoglyph of ``7'' in Figure\ref{fig:homoglyphs7}.

\begin{figure}[H]
    \centering
    \includegraphics[width=0.95\textwidth]{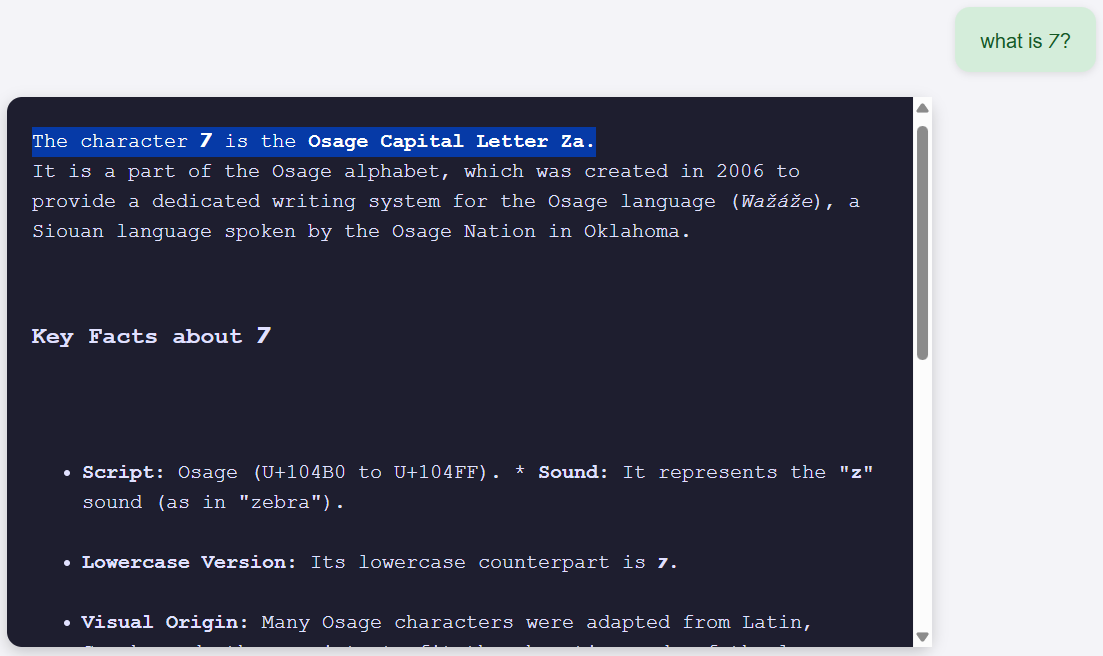}
    \caption{Illustrating the unrecognizability of a homoglyph of numeric character ``7''. This particular example is a response from Gemini.}
    \label{fig:gemini_response_for_7}
\end{figure}

Since our goal is not just to fool the LLMs but also keep the question easily readable by humans we need to be careful about the choice of homoglyph that will be used for the perturbation. Hence to find out an effective homoglyph we select a homoglyph that is easily readable by humans and not recognized by the LLMs. To confirm that the LLMs do not understand the homoglyph, we ask the LLMs about it and if the LLM can not recognize it then we use that to perturb the questions. For example, if we ask about the second homoglyph of ``7'', most popular LLMs are unable to recognize it although it is quite easily readable to humans (see Figure~\ref{fig:gemini_response_for_7}).

\subsection{Step 4: Perturb the Question}

\begin{figure}[H]
    \centering
    \includegraphics[width=0.95\textwidth]{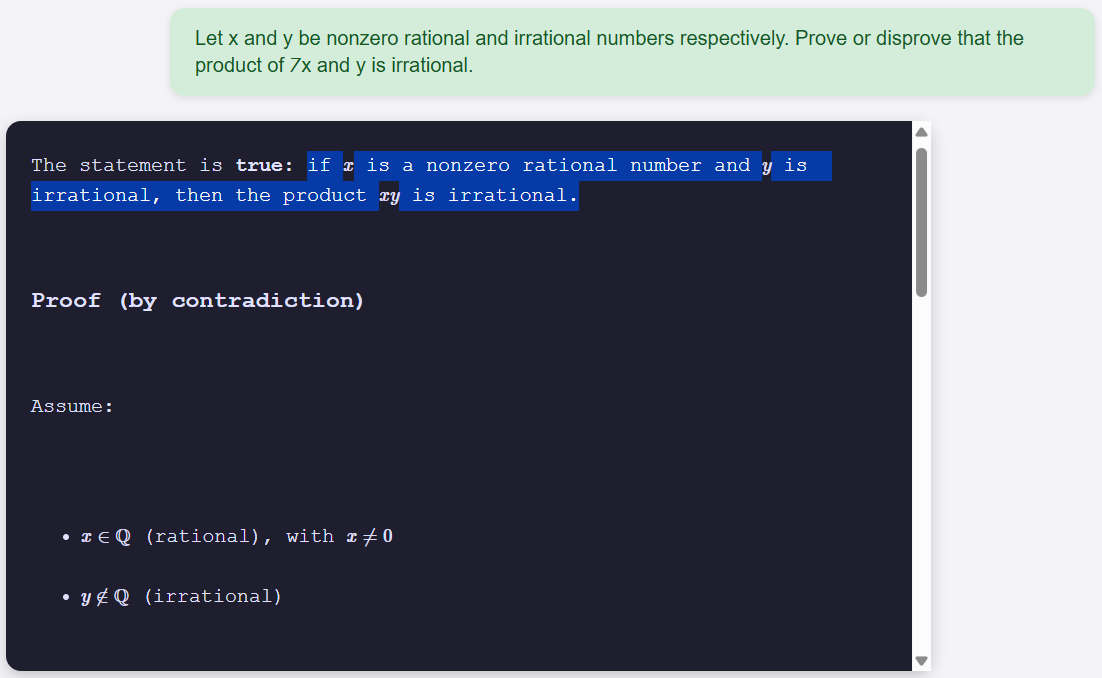}
    \caption{Illustrating an example where we added a perturbation of ``7'' as the coefficient of ``x''. The model does not recognize the perturbed character and uses ``x'' instead of ``7x'' since it has seen the usage of ``x'' in this context in the dataset based on which the model is trained on. This particular example is a response from GhatGPT.}
    \label{fig:irrational_chatgpt}
\end{figure}

The last step of our method is to perturb the question we get from Step 2 by a homoglyph we get from Step 3. For example, we take the question ``Let $x$ and $y$ be nonzero rational and irrational numbers respectively. Prove or disprove that the product of $7x$ and $y$ is irrational.''. And we replace the coefficient ``7'' by the homoglyph shown in Figure~\ref{fig:gemini_response_for_7}. And due to the perturbation the LLM provides an incorrect result as shown in Figure~\ref{fig:irrational_chatgpt}. The LLM does not recognize the coefficient seven as it is a perturbation of the number. The model without hesitation replaces ``7x'' by ``x''; which is a clear indication that the models are biased towards the original questions as they appear in the popular discrete mathematics textbooks and in many online websites, and these contents are used for pretraining the model.

We want to emphasize here that we modify the question twice in Step 2. For example, let the original question be ``Prove or disprove that the product of a nonzero rational number and an irrational number is irrational.'' We first modify it to ``Let $x$ and $y$ be nonzero rational and irrational numbers respectively. Prove or disprove that the product of $x$ and $y$ is irrational.'' in order to introduce mathematical expressions to perturb. Then we further modify the question to ``Let $x$ and $y$ be nonzero rational and irrational numbers respectively. Prove or disprove that the product of $7x$ and $y$ is irrational.'' and introduce the coefficient ``$7$'' to ``$x$''. Now the first modification might not be necessary since the original question can already contain some mathematical expression. However, the second modification is needed because we want to use the fact that the models are biased towards the training dataset and we want to ask them something that is not in the training dataset.

\section{Evaluation}\label{sec:evaluation}

\begin{figure}[H]
    \centering
    \includegraphics[width=0.65\textwidth]{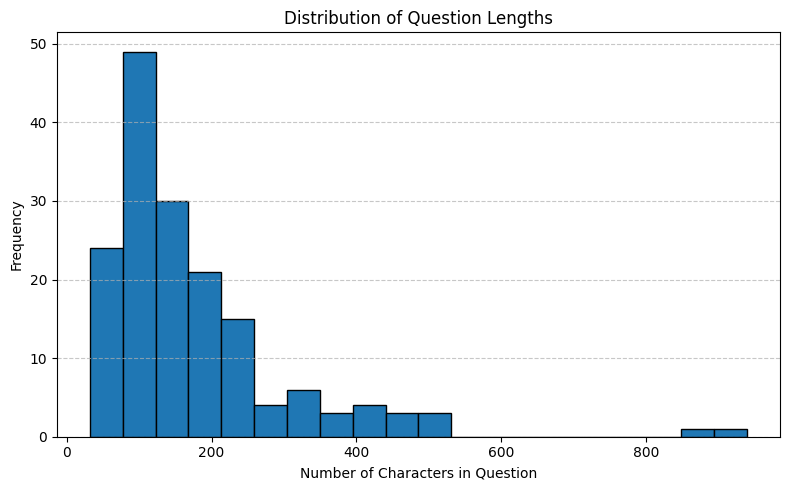}
    \caption{Illustrating the distribution of the number of characters across the 164 questions considered.}
    \label{fig:questions_chars}
\end{figure}

We consider a collection of 12 homework assignments from a standard discrete mathematics course. Each homework contains approximately 15 questions, covering a range of topics typically included in such courses, such as logic, proofs, number theory, and combinatorics. In total, the dataset consists of 164 questions. A significant portion of these questions are drawn from the widely used textbook Discrete Mathematics and Its Applications by Kenneth H. Rosen~\cite{discrete2007mathematics}, ensuring that the problems are representative of conventional coursework in the field. Out of our 164 questions, the minimum number of characters is 32. The maximum and mean are 939 and 173.45, respectively. The standard deviation is 134.93 (see Figure~\ref{fig:questions_chars}).

Some of the questions in these homework assignments include images, diagrams, or other visual components that are necessary for fully understanding or solving the problem. However, in this study, we restrict our focus exclusively to text-based questions. This decision is motivated by the scope of our experiments, which are designed to operate on textual input only and do not incorporate image processing capabilities.

After filtering out questions that contain any visual elements, we are left with a subset of 69 purely text-based questions. These questions form the basis of our experimental evaluation. The remaining questions, which rely on images or diagrams, are excluded from the analysis to ensure consistency and to avoid introducing ambiguity or incomplete information into the dataset.

For our experiments, we have used some of the most popular LLMs such as ChatGPT, Gemini, and Claude. We have tested our method on a few questions using our method, and the initial experiments have shown effective to all these LLMs. However, to try a prompt in Claude one needs to sign in with an account. Which means all the prompts we are going to try will be saved on behalf of that account. There is a possibility that the effectiveness of our method may get reduced if all the history is saved to the system. Hence, for the broader experiment on 69 questions, we only used ChatGPT and Gemini.

\begin{figure}[H]
    \centering
    \begin{subfigure}{0.48\textwidth}
        \centering
        \includegraphics[width=\linewidth]{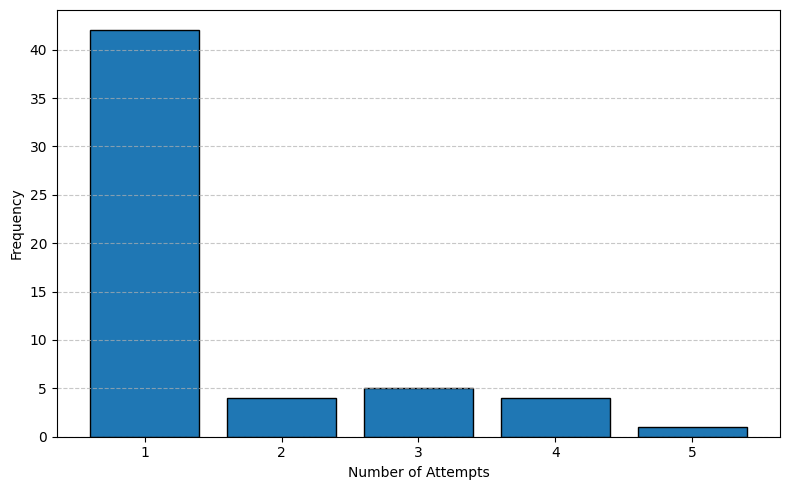}
        \caption{ChatGPT}
        \label{fig:ChatGPT}
    \end{subfigure}
    \hfill
    \begin{subfigure}{0.48\textwidth}
        \centering
        \includegraphics[width=\linewidth]{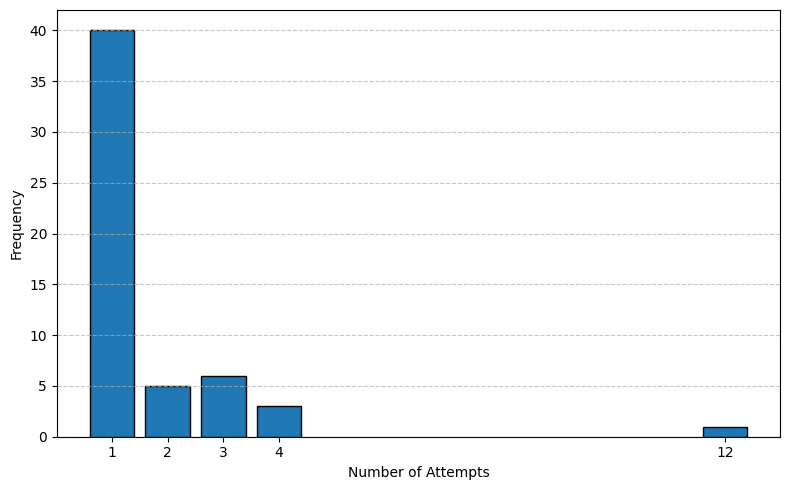}
        \caption{Gemini}
        \label{fig:Gemini}
    \end{subfigure}
    \caption{Illustrating the distribution of the number of attempts to fool the models across the 164 questions considered.}
    \label{fig:attempts}
\end{figure}

\subsection{Experimental Result}

For each question we need only a few number of attempts. For example, for ChatGPT the minimum, maximum, mean, and standard deviation are 1, 5, 1.54, and 1.03 respectively. For Gemini the minimum, maximum, mean, and standard deviation are 1, 12, 1.67, and 1.66 respectively (see Figure~\ref{fig:attempts}).

\begin{figure}[H]
    \centering
    \begin{subfigure}{0.48\textwidth}
        \centering
        \includegraphics[width=\linewidth]{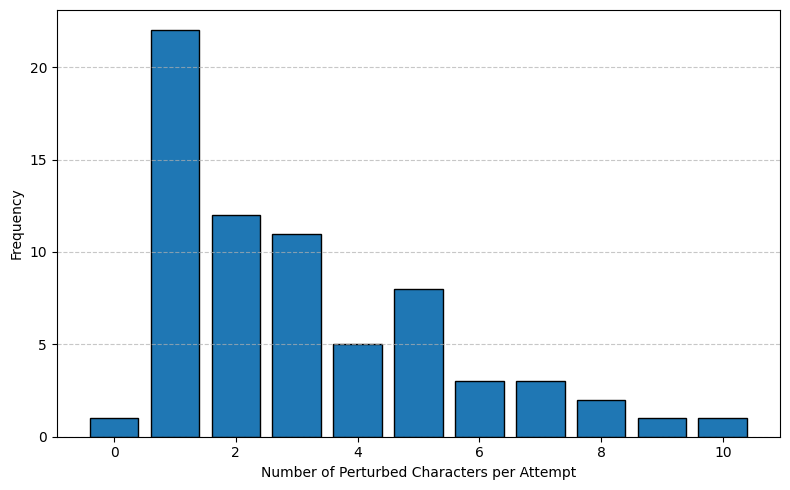}
        \caption{ChatGPT}
        \label{fig:ChatGPTnPerturbs}
    \end{subfigure}
    \hfill
    \begin{subfigure}{0.48\textwidth}
        \centering
        \includegraphics[width=\linewidth]{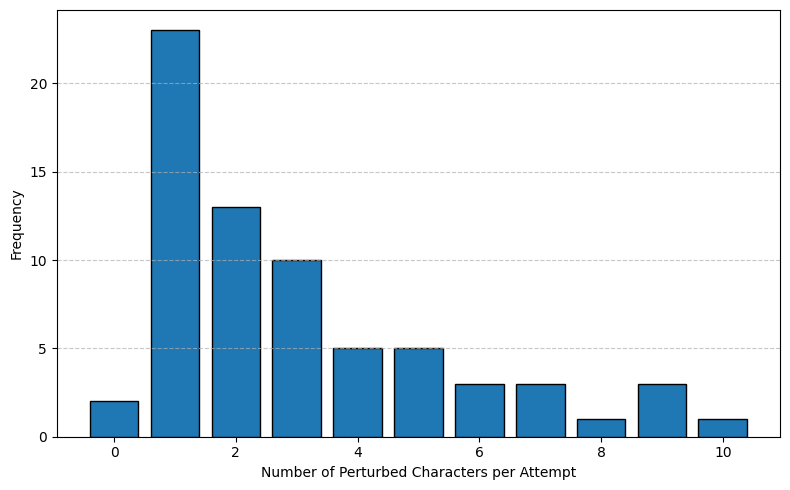}
        \caption{Gemini}
        \label{fig:GemininPerturbs}
    \end{subfigure}
    \caption{Illustrating the distribution of the number of perturbed characters to fool the models across the 164 questions considered.}
    \label{fig:chars}
\end{figure}

Also, for each attempt we need only a few number of characters to perturb. For example, for ChatGPT the minimum, maximum, mean, and standard deviation are 0, 10, 3.09, and 2.29 respectively. For Gemini the minimum, maximum, mean, and standard deviation are 0, 10, 3.01, and 2.44 respectively (see Figure~\ref{fig:chars}).

\begin{figure}[H]
    \centering
    \includegraphics[width=0.95\textwidth]{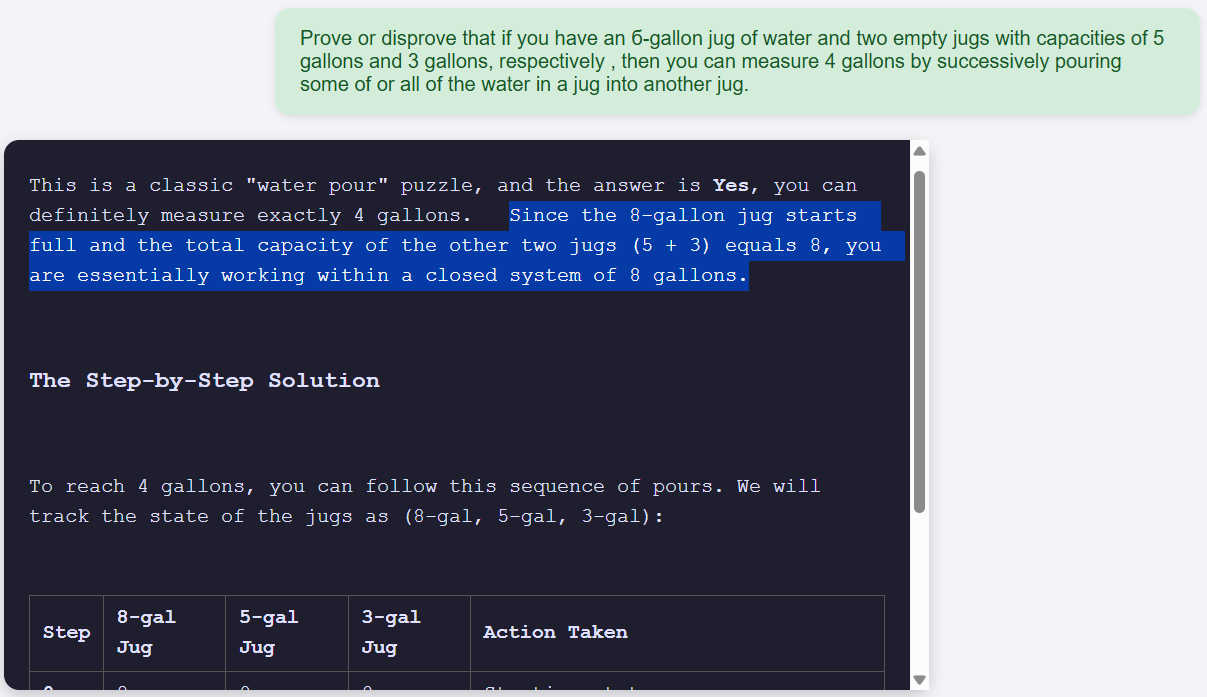}
    \caption{Illustrating the unrecognizability of a homoglyph of numeric character ``6''. This particular example is a response from Gemini.}
    \label{fig:gallon_gemini}
\end{figure}

In most of our experiments we are able to fool the models just by perturbing one character. For example, consider the classic discrete mathematics questions ``Prove or disprove that if you have an 8-gallon jug of water and two empty jugs with capacities of 5 gallons and 3 gallons, respectively , then you can measure 4 gallons by successively pouring some of or all of the water in a jug into another jug.'' 
If we replace ``8'' by the perturbed character of ``6'', then the model still assumes that we are asking about ``8-gallon'' although any human will read it as ``6-gallon'', see Figure~\ref{fig:gallon_gemini}. For example, Gemini will start with an 8-gallon jug despite the question asking about a six gallon jug. The reason is the textbook (as well as all online blogs) uses an 8-gallon jug and the six in the question is not a regular six, it is a character that represents something different but the models are not aware that it looks like six. Using this technique we are able to always fool the models just changing one character.

\begin{figure}[H]
    \centering
    \includegraphics[width=0.95\textwidth]{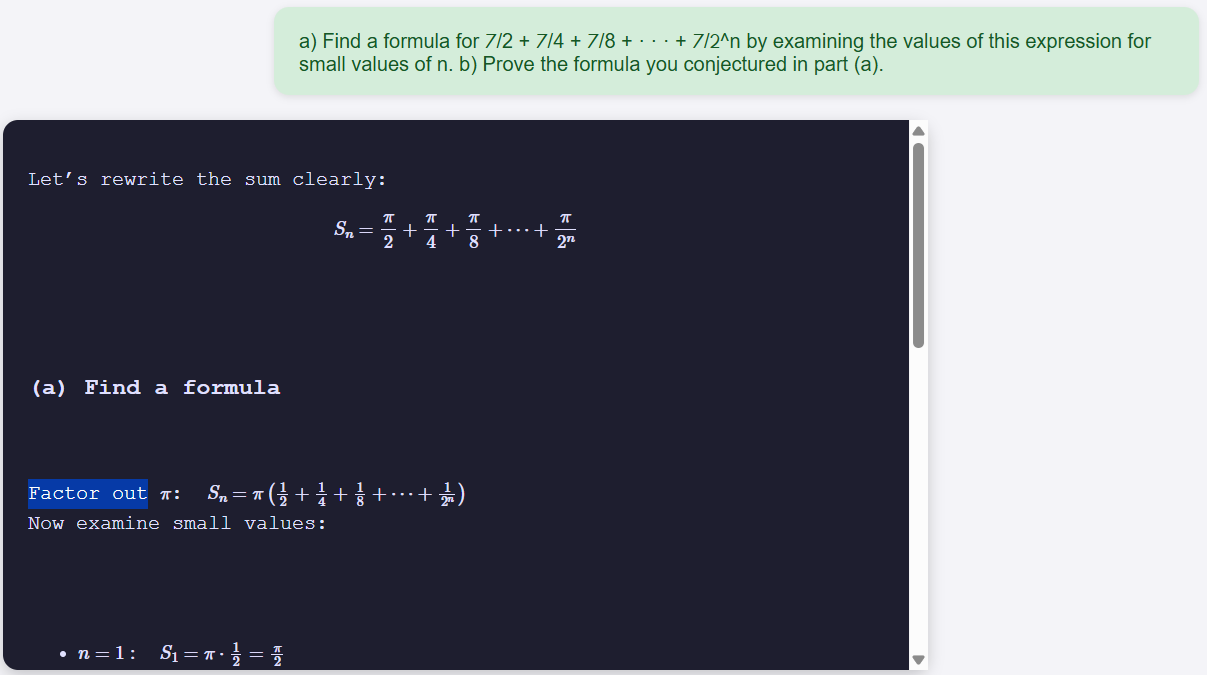}
    \caption{Illustrating the observation that the more perturbations can become less effective. Here we perturb four characters, all of them are homoglyphs of ``7''. Although the model considered the homoglyph as $\pi$, however, it can now take that as a common factor and do the rest of the analysis, see Figure~\ref{fig:factor_chat2} where we can fool the model by just perturbing one character. This particular example is a response from ChatGPT.}
    \label{fig:factor_chat}
\end{figure}

\begin{figure}[H]
    \centering
    \includegraphics[width=0.95\textwidth]{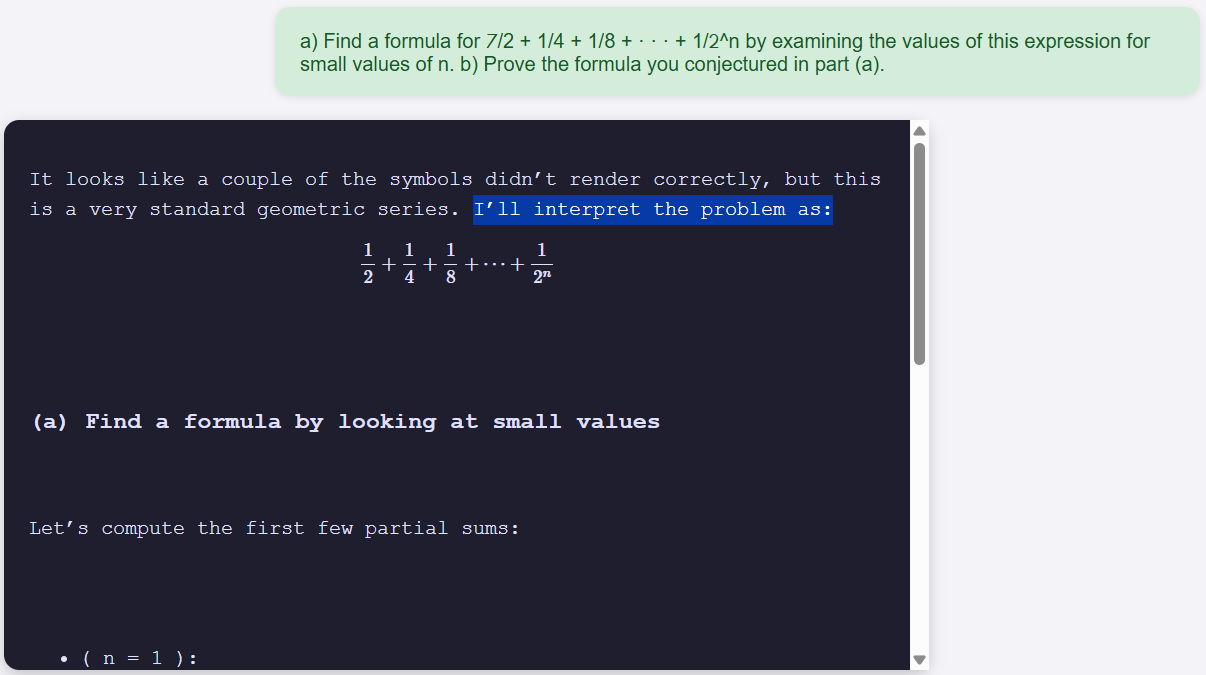}
    \caption{Illustrating the observation that less perturbations can become more effective. Here we perturb only one character, it is the homoglyphs of ``7''. The model considered the homoglyph as $1$, and solved a different problem. This particular example is a response from ChatGPT.}
    \label{fig:factor_chat2}
\end{figure}

Not only is the number of perturbations required is small to fool the model, but also we have noticed that more perturbation can make the process less effective. For example, consider the question ``a) Find a formula for $1/2 + 1/4 + 1/8 + · · · + 1/2^n$ by examining the values of this expression for small values of n. b) Prove the formula you conjectured in part (a).'' One possible way to perturb the question is to change all ``1'' by an effective homoglyph of ``7'', see Figure~\ref{fig:factor_chat}. Although the model considered the homoglyph as $\pi$, however, it can now take that as a common factor and do the rest of the analysis, see Figure~\ref{fig:factor_chat2} where we can fool the model by just perturbing one character.

\begin{figure}[H]
    \centering
    \includegraphics[width=0.95\textwidth]{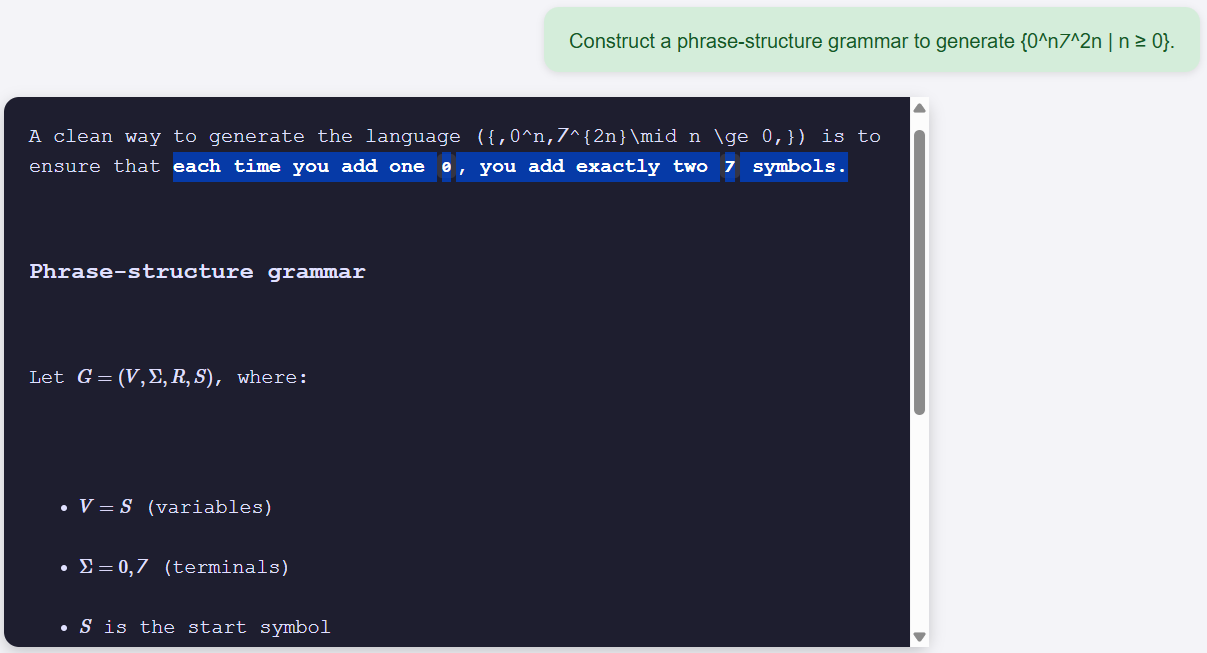}
    \caption{Illustrating the observation that not always perturbing numeric characters are effective; in this question the numeric character ``7'' just represents a symbol. To effectively perturb we need to select characters that are involved with arithmetic, see Figure~\ref{fig:grammar_chat2}. This particular example is a response from ChatGPT.}
    \label{fig:grammar_chat}
\end{figure}

\begin{figure}[H]
    \centering
    \includegraphics[width=0.95\textwidth]{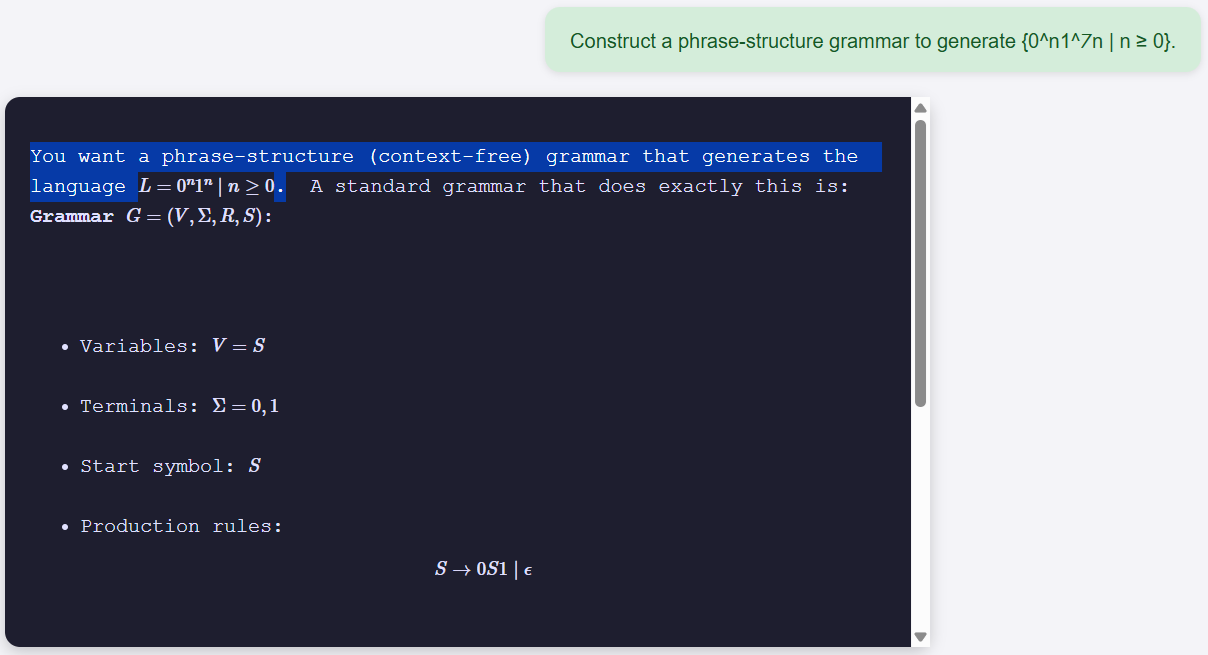}
    \caption{Illustrating the observation that perturbing numeric characters are effective. In this question we perturb the numeric character ``7'' which is involved with arithmetic. This particular example is a response from ChatGPT.}
    \label{fig:grammar_chat2}
\end{figure}

In our experiments we have observed that perturbing numeric characters are effective, however there exists questions where the numeric characters just represent different symbols. For example, consider the question ``Construct a phrase-structure grammar to generate $\{0^n 1^{2n} | n \geq 0\}$.'' Here the characters ``0'' and ``1'' are just representing some symbols and if we perturb them, then it will be ineffective, see Figure~\ref{fig:grammar_chat}. On the other hand, ``2'' is a numeric character and if we replace it by a homoglyph of ``7'', then it will be an effective perturbation, see Figure~\ref{fig:grammar_chat2}.

\subsection{An Interactive Tool}

\begin{figure}[H]
    \centering
    
    \begin{subfigure}{0.5\textwidth}
        \centering
        \includegraphics[width=\textwidth]{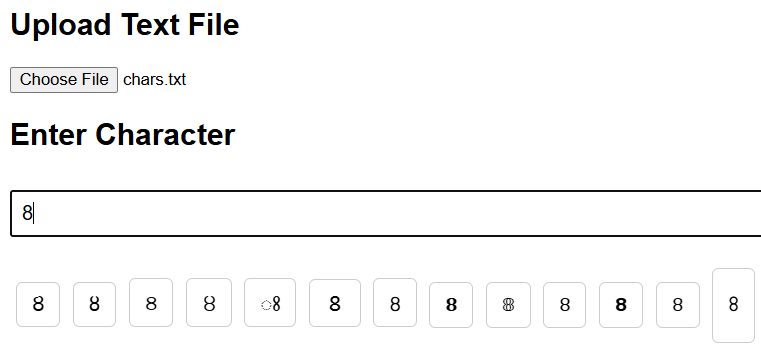}
        \caption{There is a text box to enter a character. Once the character is entered the user will be able to select any homoglyphs available in the file uploaded for that character.}
    \end{subfigure}
    
    \vspace{0.5cm}
    
    \begin{subfigure}{0.2\textwidth}
        \centering
        \includegraphics[width=\textwidth]{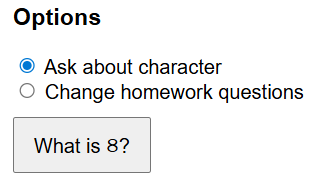}
        \caption{ One type of prompt is to figure out whether the LLM identifies the homoglyph, e.g. ``What is 8?''}
    \end{subfigure}
    
    \vspace{0.5cm}
    
    \begin{subfigure}{0.5\textwidth}
        \centering
        \includegraphics[width=\textwidth]{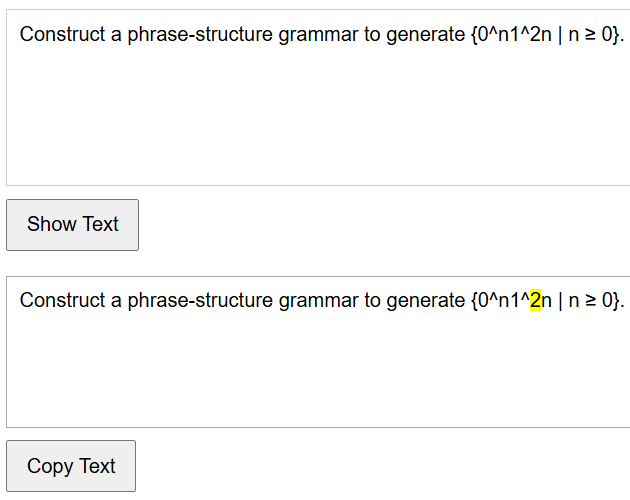}
        \caption{The user can perturb a question using all the homoglyphs available. There is a text area to paste the original question. The user can click on the ``Show text'' button and the question will be shown in a division below. The user then can select any character, the character will get highlighted. The user can select any homoglyph and the highlighted character will get replaced by that homoglyph. There is also a button ``Copy text'', if the user clicks that button then the perturbed question will get copied in the clipboard.}
    \end{subfigure}
    
    \caption{Illustrating the interactive tool.}
    \label{fig:tool}
\end{figure}

We have also developed an interactive tool where we can apply our method efficiently, see Figure~\ref{fig:tool}. The user will first upload the text file containing the homoglyphs. There is a text box to enter a character. Once the character is entered the user will be able to select any homoglyphs available in the file uploaded for that character. The tool provides options to generate two types of prompts. One type of prompt is to figure out whether the LLM identifies the homoglyph, e.g. ``What is 8?'' If the user clicks on the prompt it will be automatically copied in the clipboard and the user can paste it to any LLM. Also, the user can perturb a question using all the homoglyphs available. There is a text area to paste the original question. The user can click on the ``Show text'' button and the question will be shown in a division below. The user then can select any character, the character will get highlighted. The user can select any homoglyph and the highlighted character will get replaced by that homoglyph. There is also a button ``Copy text'', if the user clicks that button then the perturbed question will get copied in the clipboard. Then the user can use it to any LLM.

\section{Conclusions}
We have studied the homoglyph-based perturbation techniques to ensure the learning objectives of a course. We have proposed a simple method that can effectively fool all modern LLMs. The effectiveness of our method has been shown on a set of theory problems from an introductory computer science course. We have also proposed a tool to apply our method.

\bibliographystyle{plain}
\bibliography{refs}

@inproceedings{chen2024plagiarism,
  title="Plagiarism in the age of generative {AI}: cheating method change and learning loss in an intro to {CS} course",
  author={Chen, Binglin and Lewis, Colleen M and West, Matthew and Zilles, Craig},
  booktitle={Proceedings of the Eleventh ACM Conference on Learning@ Scale},
  pages={75--85},
  year={2024}
}

@article{santos2026llm,
  title={LLM Use, Cheating, and Academic Integrity in Software Engineering Education},
  author={Santos, Ronnie de Souza and Santos, Italo and Bento, Mariana and Destefanis, Giuseppe and Magalh{\~a}es, Cleyton and Wessel, Mairieli},
  journal={arXiv preprint arXiv:2603.17060},
  year={2026}
}

@article{puthumanaillam2025lazy,
  title={The Lazy Student’s Dream: ChatGPT Passing an Engineering Course on Its Own},
  author={Puthumanaillam, Gokul and Bretl, Timothy and Ornik, Melkior},
  journal={IFAC-PapersOnLine},
  volume={59},
  number={7},
  pages={213--218},
  year={2025},
  publisher={Elsevier}
}

@inproceedings{fang2024large,
  title={Large language models are neurosymbolic reasoners},
  author={Fang, Meng and Deng, Shilong and Zhang, Yudi and Shi, Zijing and Chen, Ling and Pechenizkiy, Mykola and Wang, Jun},
  booktitle={Proceedings of the AAAI conference on artificial intelligence},
  volume={38},
  number={16},
  pages={17985--17993},
  year={2024}
}

@article{hao2025investigation,
  title={An Investigation of Robustness of LLMs in Mathematical Reasoning: Benchmarking with Mathematically-Equivalent Transformation of Advanced Mathematical Problems},
  author={Hao, Yuren and Wan, Xiang and Zhai, Chengxiang},
  journal={arXiv preprint arXiv:2508.08833},
  year={2025}
}

@article{huang2025math,
  title={MATH-Perturb: Benchmarking LLMs' Math Reasoning Abilities against Hard Perturbations},
  author={Huang, Kaixuan and Guo, Jiacheng and Li, Zihao and Ji, Xiang and Ge, Jiawei and Li, Wenzhe and Guo, Yingqing and Cai, Tianle and Yuan, Hui and Wang, Runzhe and others},
  journal={arXiv preprint arXiv:2502.06453},
  year={2025}
}

@article{colelough2025neuro,
  title={Neuro-symbolic AI in 2024: A systematic review},
  author={Colelough, Brandon C and Regli, William},
  journal={arXiv preprint arXiv:2501.05435},
  year={2025}
}

@article{discrete2007mathematics,
  title={Mathematics and its Applications},
  author={Discrete, Kenneth H Rosen},
  journal={Higher Education. 4th edition. McGraw-Hill},
  year={2007}
}

@inproceedings{salim2024impeding,
  title="Impeding {LLM}-assisted Cheating in Introductory Programming Assignments via Adversarial Perturbation",
  author={Salim, Saiful and Yang, Rubin and Cooper, Alexander and Ray, Suryashree and Debray, Saumya and Rahaman, Sazzadur},
  booktitle={Proceedings of the 2024 Conference on Empirical Methods in Natural Language Processing},
  pages={445--463},
  year={2024}
}

@inproceedings{liu2007fighting,
  title={Fighting unicode-obfuscated spam},
  author={Liu, Changwei and Stamm, Sid},
  booktitle="Proceedings of the anti-phishing working groups 2nd annual e{C}rime researchers summit",
  pages={45--59},
  year={2007}
}

@inproceedings{sokolov2020visual,
  title={Visual spoofing in content-based spam detection},
  author={Sokolov, Mark and Olufowobi, Kehinde and Herndon, Nic},
  booktitle={13th International Conference on Security of Information and Networks},
  pages={1--5},
  year={2020}
}

@article{gabrilovich2002homograph,
  title={The homograph attack},
  author={Gabrilovich, Evgeniy and Gontmakher, Alex},
  journal={Communications of the ACM},
  volume={45},
  number={2},
  pages={128},
  year={2002},
  publisher={ACM New York, NY, USA}
}

@inproceedings{creo2025silverspeak,
  title="{S}ilver{S}peak: Evading {AI}-Generated Text Detectors using Homoglyphs",
  author={Creo, Aldan and Pudasaini, Shushanta},
  booktitle="Proceedings of the 1st Workshop on {GenAI} Content Detection (GenAIDetect)",
  pages={1--46},
  year={2025}
}

@inproceedings{wolff2020attacking,
    author = "Wolff, Max and Wolff, Stuart",
    title = "Attacking neural text detectors",
    booktitle = "ICLR 2020 Workshop on Trustwory Machine Learning",
    year = 2020,
}

@inproceedings{cooper2023hiding,
  title={Hiding in plain sight: Tweets with hate speech masked by homoglyphs},
  author={Cooper, Portia and Surdeanu, Mihai and Blanco, Eduardo},
  booktitle={Findings of the Association for Computational Linguistics: EMNLP 2023},
  pages={2922--2929},
  year={2023}
}

@article{pudasaini2025survey,
  title={Survey on AI-Generated Plagiarism Detection: The Impact of Large Language Models on Academic Integrity: S. Pudasaini et al.},
  author={Pudasaini, Shushanta and Miralles-Pechu{\'a}n, Luis and Lillis, David and Llorens Salvador, Marisa},
  journal={Journal of Academic Ethics},
  volume={23},
  number={3},
  pages={1137--1170},
  year={2025},
  publisher={Springer}
}

@inproceedings{adnan2025cheating,
  title={Cheating Using AI and Copy-Pasting from LLMs: New Realities in Higher Education},
  author={Adnan, Airil Haimi Mohd and Salim, Mohamad Safwat Ashahri Mohd and Shah, Dianna Suzieanna Mohamad and Yusuf, Asmahanim Haji Mohamad and Salim, Mohd Nur Fitri Mohd and Tahir, Mohd Haniff Mohd},
  booktitle={International Conference on Business and Technology},
  pages={399--410},
  year={2025},
  organization={Springer}
}

\end{document}